\documentclass[iop,apj,appendixfloats]{emulateapj}
\usepackage{apjfonts}
\usepackage{graphicx}
\usepackage{amsmath}
\usepackage{amssymb}
\usepackage{color}
\usepackage{url}
\usepackage{hyperref}

\gdef\xx[#1]{\textcolor{red}{#1}}

\gdef\msun{M$_{\odot}$}

\gdef\gal{NGC\,5907}
\gdef\ma{mag\,arcsec$^{-2}$}

\newcommand{\GG}[1]{}

\lefthead{van Dokkum et al.}
\righthead{}
\slugcomment{Accepted for publication in ApJ Letters}
\begin{document}

\newcommand\XXX[1]{{\textcolor{red}{\textbf{x\ #1\ x}}}}

\title{Dragonfly imaging of the galaxy NGC\,5907: a different view of the iconic stellar stream}


\author{Pieter van Dokkum\altaffilmark{1},
Colleen Gilhuly\altaffilmark{2},
Ana Bonaca\altaffilmark{3},
Allison Merritt\altaffilmark{4},
Shany Danieli\altaffilmark{1,5},
Deborah Lokhorst\altaffilmark{2},
Roberto Abraham\altaffilmark{2},
Charlie Conroy\altaffilmark{3},
Johnny P.\ Greco\altaffilmark{6}
\vspace{8pt}}

\altaffiltext{1}
{Astronomy Department, Yale University, 52 Hillhouse Ave,
New Haven, CT 06511, USA}
\altaffiltext{2}
{Department of Astronomy \& Astrophysics,
University of Toronto, 50 St.\ George
Street, Toronto, ON M5S 3H4, Canada}
\altaffiltext{3}
{Harvard-Smithsonian Center for Astrophysics, 60 Garden Street,
Cambridge, MA, USA}
\altaffiltext{4}
{Max-Planck-Institut f\"ur Astronomie,
K\"onigstuhl 17, D-69117 Heidelberg, Germany}
\altaffiltext{5}
{Physics Department, Yale University, 52 Hillhouse Ave,
New Haven, CT 06511, USA}
\altaffiltext{6}
{Center for Cosmology and AstroParticle Physics (CCAPP), The Ohio State University, Columbus, OH 43210, USA}

\begin{abstract}

In 2008 it was reported that the stellar stream of the
edge-on spiral \gal\
loops twice around the galaxy, enveloping it
in a giant corkscrew-like
structure.  Here we present imaging of this iconic object
with the Dragonfly Telephoto Array, reaching a $1\sigma$ surface brightness
level of $\mu_g= 30.3$\,\ma\ on spatial scales of $1\arcmin$ (the
approximate width of the stream).
We find a qualitatively
different morphology from that
reported in the 2008 study. The Dragonfly data 
do not show two loops but
a single curved stream with a total length of $45\arcmin$ (220\,kpc).
The surface brightness of the stream ranges from
$\mu_g \approx 27.6$\,\ma\ to $\mu_g\approx 28.8$\,\ma, and it
extends significantly beyond the region where tidal features had
previously been detected.
We find a density enhancement near the luminosity-weighted
midpoint of the stream which we identify as the likely remnant
of a nearly-disrupted progenitor galaxy. A restricted N-body
simulation provides a qualitative match to the detected features.
In terms of its spatial extent and stellar mass the stream is
similar to Sagittarius, and
our results demonstrate the
efficacy of low surface brightness-optimized telescopes
for obtaining maps of such large streams 
outside the Local Group.
The census of these rare, relatively high mass events
complements the census
of common, low mass ones that is provided by studies of streams
in the Milky Way halo.


\end{abstract}


\section{Introduction}

Stellar streams, the debris of tidally-disrupted globular clusters or galaxies, provide
information on the frequency of the accretion of small objects onto larger ones 
(see, e.g., {Bullock} \& {Johnston} 2005). As their morphologies reflect their orbits 
they are also probes of the gravitational potential, and they have
been used as a tool to constrain
the mass and structure of dark matter halos ({Moore} {et~al.} 1999; {Ibata} {et~al.} 2002; {Helmi} 2004; {Law} \& {Majewski} 2010; {Bovy}, {Erkal}, \& {Sanders} 2017; {Bonaca} \& {Hogg} 2018).

In the Milky Way dozens of stellar streams have been identified
(see {Grillmair} \& {Carlin} 2016; {Shipp} {et~al.} 2018), with
Sagittarius ({Ibata} {et~al.} 1997), Palomar 5 ({Odenkirchen} {et~al.} 2001),
Monoceros ({Newberg} {et~al.} 2002), and the ``orphan stream'' ({Belokurov} {et~al.} 2007) among the most prominent
examples. The number of confirmed and candidate streams is increasing rapidly, 
thanks
to the increased contrast attainable with Gaia and deep star count maps
(see, e.g., {Malhan}, {Ibata}, \& {Martin} 2018; {Bonaca} {et~al.} 2019). Likewise, M31 is
home to many tidally-disrupting satellite objects,
ranging from low mass ``stretched'' objects such as Andromeda XIX ({McConnachie} {et~al.} 2008)
to the major event, or events, that were responsible for shaping the complex structure of the M31
halo ({D'Souza} \& {Bell} 2018, and references therein).

At distances $D\gtrsim 5$\,Mpc streams can be identified by
the smooth integrated light of their stellar populations
(e.g., {Arp} 1966; {Malin} \& {Hadley} 1997; {Mihos} {et~al.} 2005; {van Dokkum} 2005; {Bell} {et~al.} 2006; {Mart{\'{\i}}nez-Delgado} {et~al.} 2010; {Atkinson}, {Abraham}, \&  {Ferguson} 2013).
Such integrated-light measurements typically
do not reach the same stellar density limits as star count surveys but probe a much greater
volume of the Universe
(see {Danieli}, {van Dokkum}, \&  {Conroy} 2018, for a quantitative discussion). The combination of studies of frequent,
low mass accretion events in the Local Group with systematic integrated-light
surveys of rare, high mass events around other galaxies should ultimately provide a complete census
of present-day accretion-driven galaxy growth.

\begin{figure*}[htbp]
  \begin{center}
  \includegraphics[width=1.0\linewidth]{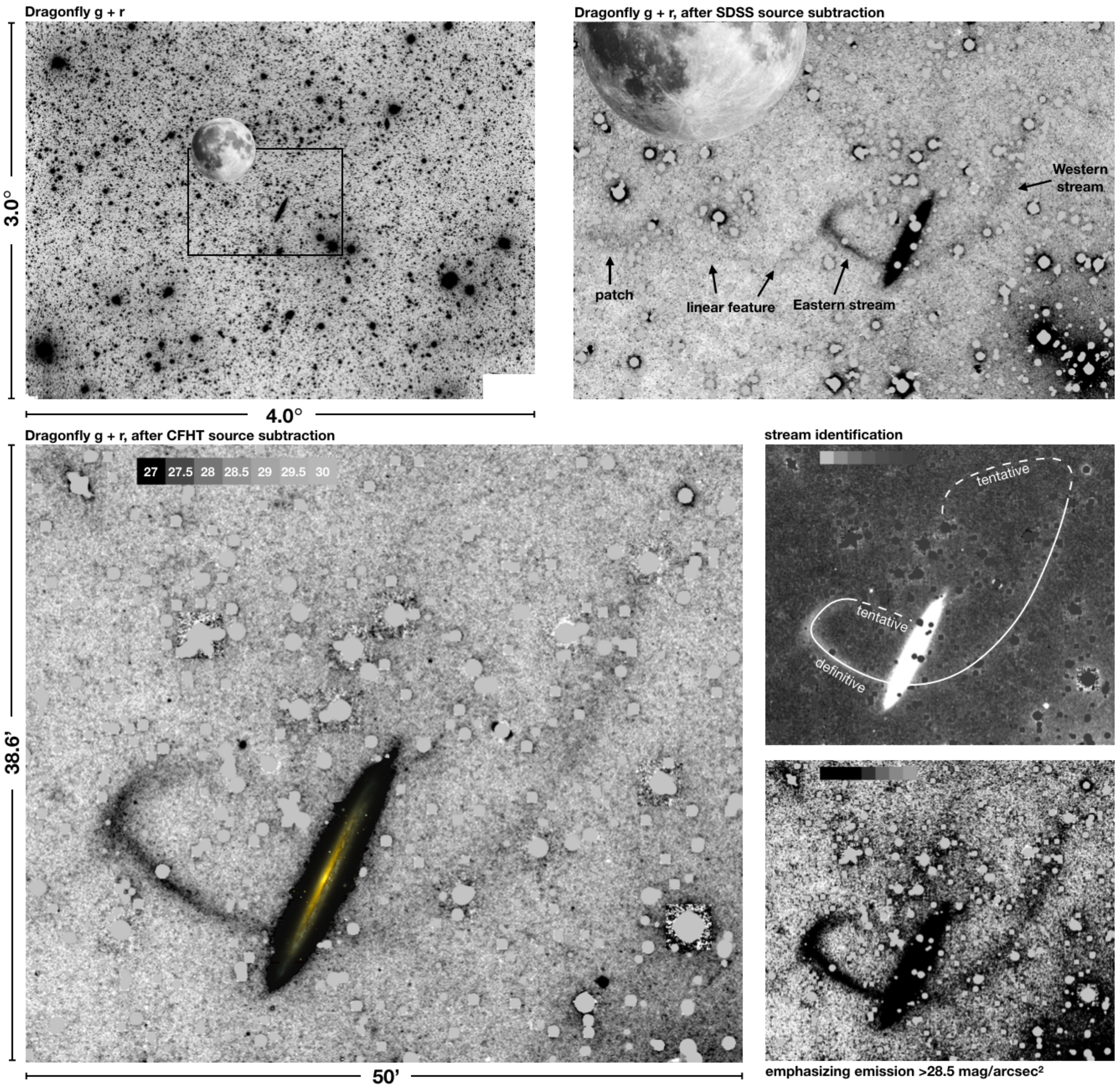}
  \end{center}
\vspace{-0.2cm}
    \caption{
Dragonfly imaging of the \gal\ field, with North up and East to the left.
{\em Top left:} sum of the $g$ and $r$ band images.  {\em Top right:} zoom on the vicinity of
\gal, after subtracting a model of compact emission in the frame.  The image shows
a single coherent stellar stream with a length of $\approx 45\arcmin$ that crosses
the galaxy.
We also identify a thin, linear feature to the East and a low surface brightness
patch $1\arcdeg$ from \gal.  {\em Bottom panels:} the region of the \gal\ 
stream, at three different scalings. The
scale bar at the top indicates the surface brightness in AB mag\,arcsec$^{-2}$.
}
\label{overview.fig}
\end{figure*}

One of the best-known tidal features outside of the Local Group
is the stellar stream associated with \gal, an edge-on spiral galaxy with a stellar mass of
$\approx 8\times 10^{10}$\,M$_{\odot}$ ({Laine} {et~al.} 2016)
at a distance of 17\,Mpc ({Tully}, {Courtois}, \& {Sorce} 2016).
The stream was discovered by {Shang} {et~al.} (1998) and {Zheng} {et~al.} (1999),
who detected sections of a loop around the disk of \gal\ using the Beijing
Astronomical Observatory 0.6/0.9\,m Schmidt telescope.
This was a remarkable discovery, as previous deep
optical and H{\sc i} searches 
had not uncovered any tidal features associated with \gal\
(see {Sancisi} 1976; {Sasaki} 1987; {Sackett} {et~al.} 1994).
The galaxy was subsequently imaged by
{Mart{\'\i}nez-Delgado} {et~al.} (2008) [hereafter M08], using a 0.5\,m Ritchey-Chr\'etien telescope.
M08 report that 
the stream exhibits not one but two complete loops, enveloping  \gal\ in a
giant corkscrew-like structure. Their evocative image, whose main
features could be
reproduced with an N-body model, has taken on an iconic status,
serving as a powerful demonstration of the shredding of a small galaxy.\footnote{We
note that {Wang} {et~al.} (2012) interpret the M08 data as evidence of a major merger.}
Some years later \gal\ was also observed by  {Laine} {et~al.} (2016), who
combined data from the  Spitzer Space Telescope with optical Subaru
images. These authors studied the part of the stream that was
detected  by {Shang} {et~al.} (1998) and do not comment on the second loop that was
reported by M08.

Here we report on new low surface brightness imaging of \gal\ over a wide
field, as part of an imaging campaign of
nearby galaxies with the Dragonfly Telephoto Array ({Abraham} \& {van Dokkum} 2014).
We are conducting two surveys, the Dragonfly Nearby Galaxies Survey
({Merritt} {et~al.} 2016) and the Dragonfly Edge-on Galaxies Survey (C.\ Gilhuly et al.,
in preparation); \gal\ was one of the first targets of the edge-on survey.


\section{Data}
\subsection{Observations and reduction}

The observations were obtained with the Dragonfly Telephoto Array, a low surface brightness-optimized
telescope consisting of 48 Canon 400\,mm f/2.8 II telephoto lenses. Its basic design
is described
in {Abraham} \& {van Dokkum} (2014), {Merritt} {et~al.} (2016), and {Zhang} {et~al.} (2018).
The current 48-lens array is described in S.~Danieli et al., in preparation. Briefly, each lens
is coupled to an SBIG STT-8300M
camera offering a $2\fdg6 \times 1\fdg9$ instantaneous field of view
with $2\farcs 8$ native pixels and
a FWHM spatial resolution of $\approx 6\farcs 7$. The lenses are intentionally offset
from one another by $\approx 10$\,\% of the field of view, giving 48 independent sightlines.
Data are taken with large ($\approx 25\arcmin$) dithers between exposures, providing
further redundancy.
As the data are sky-limited in our 600\,s integrations the telescope behaves optically like
a 1.0\,m f/0.4 refractor with superb optical surfaces and near-perfect baffling.
Twenty-four lenses are equipped with SDSS $g$ filters and 24 with SDSS $r$ filters.

The data reduction is gate-based, executing multiple quality tests on each frame as it progresses
through the pipeline. 
The background modeling is done in two stages. After the first stage a mask
is generated containing all detected
emission in the co-added image. This is used in the second stage to mask all emission
sources from the individual
raw frames prior to fitting the background with a two-dimensional
3$^{\rm d}$-order polynomial. In this step variation on scales exceeding
$\sim 0\fdg9 \times 0\fdg6$ is removed;
features that are smaller in at least one dimension
(such as the stellar stream, which has a width
of $\approx 0\fdg 02$) remain unaffected.
The pipeline is described in detail in Jielai Zhang's PhD
thesis\footnote{\url{https://jielaizhang.github.io/files/Zhang\_Jielai\_201811\_PhD\_Thesis\_excludech4.pdf}} and in S.~Danieli et al., in preparation.
The total number of frames that went into the final \gal\ stacks is 618 in $g$ and 762 in $r$;
this is the equivalent of 4.8\,hr with the full 48 lens array.
The summed $g+r$ image is shown in the top left of Fig.\ \ref{overview.fig}; owing to the
dithering it covers 12\,degree$^2$, with reduced effective exposure time near the edges
of the field.

\subsection{Multi-resolution filtering}

The Dragonfly data have excellent low surface brightness sensitivity and are essentially free
of ghosts, reflections, and other artifacts. However, they suffer from crowding due to the relatively
low spatial resolution. We subtracted compact emission sources from the
data using 
multi-resolution filtering (MRF). Details are given
in P.~van Dokkum et al., in preparation. Briefly,
a flux model is created by multiplying 
an image of higher resolution (such as archival CFHT data) by a
SExtractor ({Bertin} \& {Arnouts} 1996) object map of that image.
Any detected low surface brightness features in the
high resolution data can be removed from the model at this stage.
The model is then convolved with a kernel to match the Dragonfly resolution and
subtracted. Remaining halos around bright stars are removed following a similar process
as described in {van Dokkum}, {Abraham}, \&  {Merritt} (2014). The PSF
is modeled in a $2\farcm 0\times 2\farcm 0$ box; this is generally
sufficient but we note that the very
brightest stars have detected light at larger radii in the residual image. 

The results are shown in Fig.\ \ref{overview.fig}. For the image at top right the high resolution model
was created from SDSS $g$ and $r$ images. These are shallow but have few artifacts and enable a wide field
subtraction. The images in the bottom panels were filtered using a combination of Canada
France Hawaii Telescope (CFHT) and
Beijing-Arizona All Sky Survey (BASS; {Zou} {et~al.} 2018)
imaging. The BASS data are only used to identify and remove artifacts and missing data
in the CFHT images.
We carefully checked that no low surface brightness emission is contained in the high resolution
model.
The only low surface brightness object that we removed from the model
is a previously-uncataloged dwarf galaxy.

\begin{figure}[htbp]
  \begin{center}
  \includegraphics[width=1.0\linewidth]{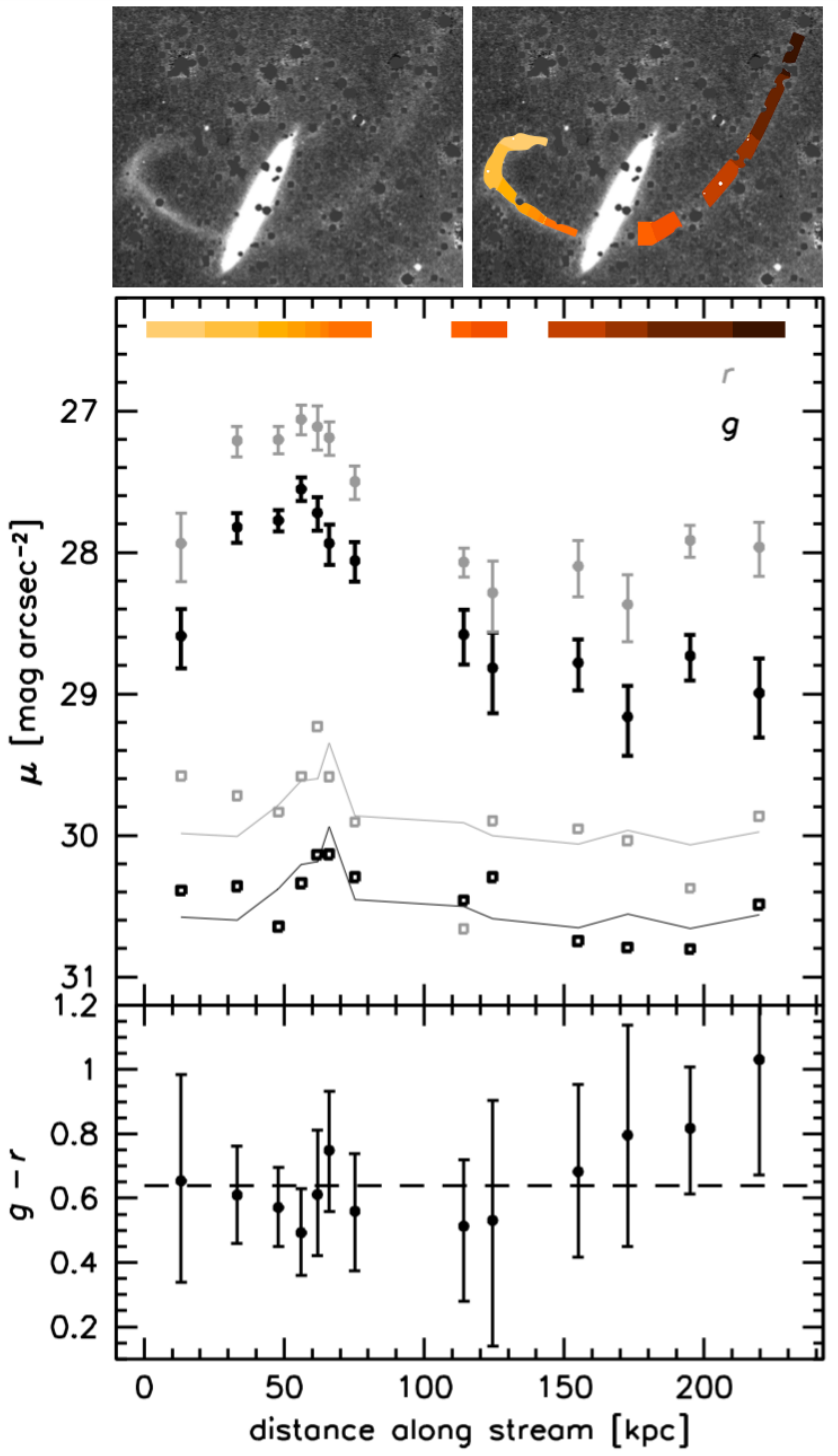}
  \end{center}
\vspace{-0.2cm}
    \caption{
Photometry along the stream. {\em Top panel:} $g$- and $r$-band surface brightness.
Open symbols indicate the $1\sigma$ uncertainty (see text).
The average
surface brightness of the stream is $\mu_g\approx 27.8$ on the
East side of the galaxy and
$\mu_g \approx 28.8$\,mag\,arcsec$^{-2}$ on the West side.
{\em Bottom panel:} $g-r$ color along the stream, with the mean
indicated by the dashed line.
}
\label{phot.fig}
\end{figure}

\section{Observational Results}

\subsection{Morphology and photometry of the stream}

\begin{figure*}[htbp]
  \begin{center}
  \includegraphics[width=0.9\linewidth]{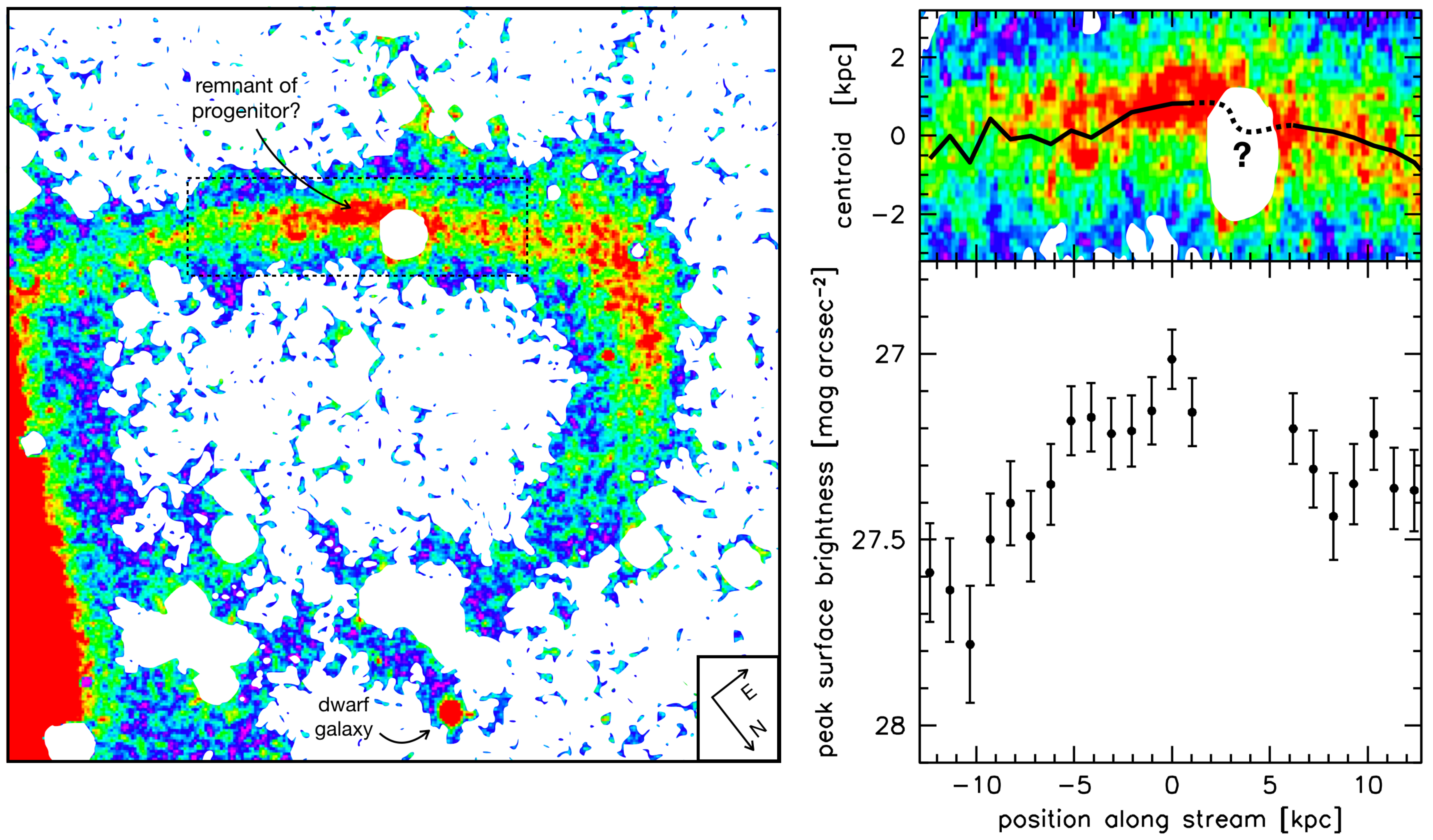}
  \end{center}
\vspace{-0.2cm}
    \caption{
{\em Left:} False color image of the Eastern stream, rotated by $142\arcdeg$.
{\em Right:} Results of Gaussian fits in $12\farcs 5$ bins along the stream segment
shown in the box at left. The top
panel shows the best-fit position and the bottom panel shows the surface brightness
of the peak of the Gaussian. There is a clear stellar density enhancement in this region,
and a possible asymmetry.
}
\label{progenitor.fig}
\end{figure*}

The Dragonfly images show a relatively straightforward stream morphology. We confirm
the existence of 
the strongly curved Eastern stream that was discovered by {Shang} {et~al.} (1998) (see top right panel of Fig.\ \ref{overview.fig}). We find that the stream continues on the West side of \gal\ at lower
surface brightness. This Western stream reaches more than twice the length of the Eastern stream.
This stream morphology is qualitatively different from the double loop
structure reported by M08; we return to this in \S\,\ref{disc.sec}.
We also detect a thin feature extending from the brightest part of the
stream to the East and
a faint patch about $1\arcdeg$ due East of \gal. These faint features are not
artifacts and are seen in
both $g$ and $r$; their nature is unclear.
Finally, we tentatively detect continuations of the stream at both
ends: there may be a thin extension of the Western stream
toward the Northeast, looping back South toward
the disk, and there is a likely continuation of the Eastern stream
toward the disk. Both these extensions are labeled ``tentative'' in
Fig.\ \ref{overview.fig}, and they are not included in our analysis.

The surface brightness along the stream in $g$ and $r$ is quantified
using aperture photometry. The apertures aim to include most of the
width of the stream. As shown in Fig.\ \ref{phot.fig} the surface brightness
reaches a peak of 
$\mu_g \approx 27.6$\,mag\,arcsec$^{-2}$ on the East side of the galaxy. On the West side
the surface brightness is lower at $\mu_g \approx 28.8$\,mag\,arcsec$^{-2}$.
The uncertainties in the data points are determined by
moving the apertures off of the actual stream and then obtaining fluxes in these
``empty'' locations. The apertures retain their position relative to each
other, with the entire set of stream apertures moved to 52 different positions.
In 13 of these positions the stream has the same orientation as the actual stream;
in the other sets of positions it is flipped in $x$, $y$, and both $x$ and $y$.
The $1\sigma$ variation in these measurements
is taken as the uncertainty (open symbols in Fig.\ \ref{phot.fig}).
These uncertainties are not constant along the stream, as they depend on the
size of the photometric aperture: for the larger apertures on the Western side the
uncertainties are smaller than for the smaller apertures on the Eastern side.
From a fit to the empirically-determined uncertainties we find that
\begin{equation}
\label{depth_g.eq}
\sigma(\mu_g)\approx{}30.25+0.5\log(A)
\end{equation}
and
\begin{equation}
\sigma(\mu_r)
\approx{}29.66+0.5\log(A),
\end{equation}
with $A$ the aperture size in arcmin$^2$
(thin lines).
The $g-r$ color along the stream is shown in the bottom panel. The data are consistent
with a constant color along the stream
of $\langle g-r\rangle\approx 0.64\pm 0.11$\,mag (where the errorbar is
the combination of $\pm 0.04$ random and $\pm 0.1$ mag
systematic uncertainty).
These results are
broadly consistent with {Shang} {et~al.} (1998) and {Laine} {et~al.} (2016), who
obtained photometry for the relatively bright Eastern part of the stream only.

The total magnitudes integrated over
all apertures are $m_g=15.5$ and
$m_r=14.8$. There are two gaps in the photometric apertures: one
coinciding with the disk and another with a bright star (see top right
panel of Fig.\ \ref{phot.fig}). Interpolating over these apertures
suggests these regions contain $\approx 10$\,\% of the light of
the stream. Assuming another 10\,\% is missed in regions that are fainter than
our detection limit,
we estimate that the total magnitudes of the stream are
$m_g\approx 15.3$ and $m_r\approx 14.6$. For $D=17$\,Mpc this
corresponds to $L_g\approx 1.8 \times 10^8$\,L$_{\odot}$.
For an analysis of the
stellar population of the stream we refer the reader to {Laine} {et~al.} (2016).

\begin{figure*}[htbp]
  \begin{center}
  \includegraphics[width=0.84\linewidth]{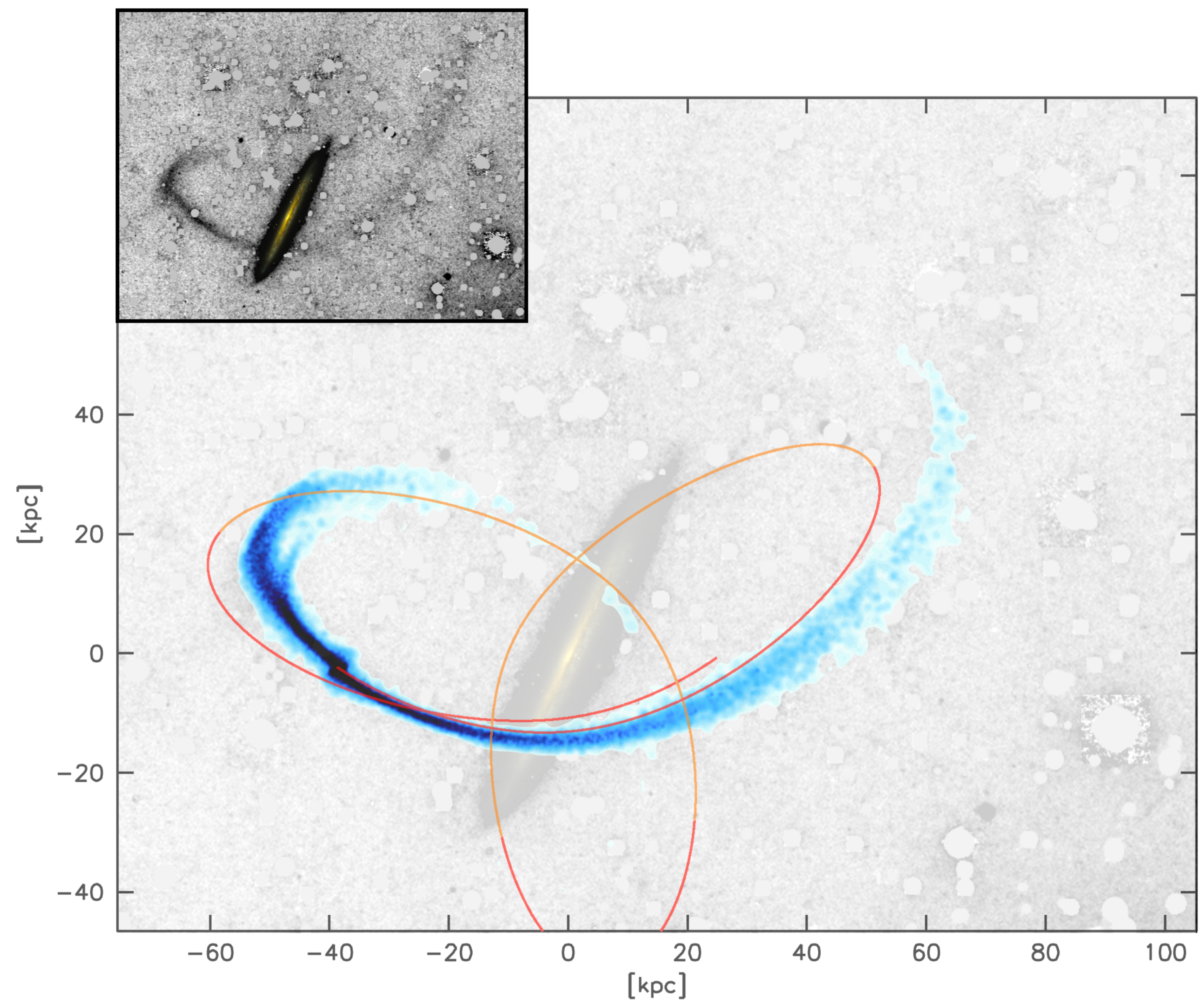}
  \end{center}
\vspace{-0.2cm}
    \caption{
Restricted
N-body simulation of a disrupting galaxy with a mass of $2\times 10^8$\,\msun, with its
present-day location matched to that of the progenitor identified in \S\,\ref{prog.sec}.
The line
indicates the most recent 2.5\,Gyr of the orbit, with red and orange alternating
every 0.5\,Gyr. 
}
\label{nbody.fig}
\end{figure*}

\subsection{Probable identification of the progenitor galaxy}
\label{prog.sec}

Stellar streams are generated by mass loss from a progenitor object along its orbit.
Generally the progenitor object is
within the densest part of the stream, is near the
luminosity-weighted midpoint of the stream, and
coincides with a displacement (as the leading and trailing streams
come from stars that became unbound at opposite Lagrange points,
toward the center and anti-center of the potential). 
These are not absolutes, as the orbital geometry, the superposition of successive passages,
and projection effects complicate
the observed morphology. 


We identify the likely remnant of the progenitor object within the region
highlighted in the left panel of
Fig.\ \ref{progenitor.fig}.
In the right panel we show
the centroid of the emission and the peak
brightness as a function of the position along this stream segment.
These values are determined by 
fitting Gaussians to the stream profile (i.e., in the 
vertical direction in Fig.\ \ref{progenitor.fig}),
averaging the $g+r$ emission in $12\farcs 5$
sections along the stream. There is a
peak in the surface brightness close to the
luminosity-weighted midpoint of the stream: $\sim 40$\,\% of the luminosity is to the
East and $\sim 60$\,\% to the West. Furthermore,
the centroid shows several $\sim 1$\,kpc-sized
offsets that could indicate the characteristic displacement of the leading and
trailing streams. A possible location is indicated with the broken line
and the question mark; unfortunately it coincides with a bright foreground star.

\begin{figure*}[htbp]
  \begin{center}
  \includegraphics[width=1.00\linewidth]{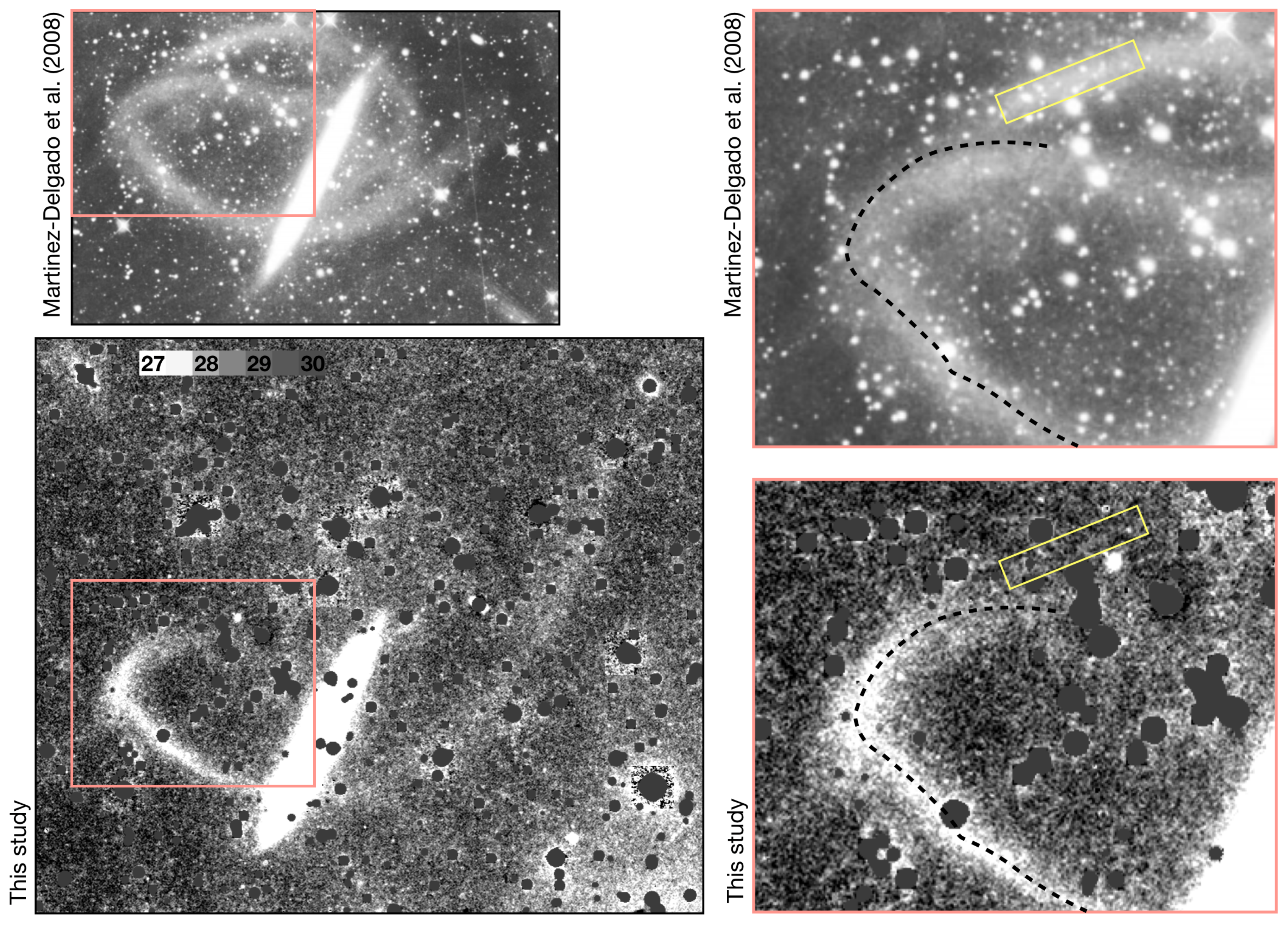}
  \end{center}
\vspace{-0.2cm}
    \caption{
Comparison of the stream morphology in M08
and in the Dragonfly $g+r$ image.
We do not confirm the
presence of a second loop.
The yellow box indicates the brightest
part of the entire M08 stream system; our limit is $\mu_g>29.4$\,\ma\
in that region.
Furthermore, the first loop is in a different location, as indicated
by the black broken line.
}
\label{comparison.fig}
\end{figure*}

\section{Dynamical stream model}

In this section we show that a tidally disrupting satellite reproduces the overall stream morphology and the identified location of the progenitor.
We followed the methodology developed for modeling streams in the Milky Way (e.g., {Price-Whelan} \& {Bonaca} 2018), and started by rotating the coordinate system such that the galaxy is aligned with the $x$-axis, $z$ is perpendicular to the disk plane and $y$ is the radial direction.
The gravitational potential is set up with the same assumptions as M08 used for the disk (mass: $8.4\times10^{10}$\,\msun, scale-length: 6.24\,kpc, scale-height: 0.26\,kpc) and bulge (mass: $2.3\times10^{10}$\,\msun, scale-radius: 0.6\,kpc). We used a more massive
halo than M08 (mass: $1.2\times10^{12}$\,\msun, scale-radius: 26\,kpc, and
$z-$axis flattening of 1.1) to better match the
recent rotation curve measurement of Posti et al.\ (2019); we tested that the M08
halo also leads to a good match to the observed stream. 

With the potential in place, we searched for the 6D location of the progenitor until we obtained an orbit that approximately matches the detected stream positions.
The progenitor is assumed to be at the approximate $x,z$ position determined in \S\,3.2, and for simplicity we set $y=0$.
The velocity is tweaked in the positive $x, z$ direction, as the morphology suggests that the Eastern stream is the leading tail.
In our model, the progenitor is currently at $\vec{x}=(-19.0,0.0,33.8)\,\rm kpc$,
$\vec{v}=(30,65,225)\,\rm km\,s^{-1}$.
Due to projection effects and the lack of kinematic data this solution is not unique, but we leave a full exploration of the parameter space to future work.

With the orbit determined, we created a mock stream using the {Fardal}, {Huang}, \& {Weinberg} (2015) method implemented in the {\tt gala}  package ({Price-Whelan} 2017).
During the most recent 2.5\,Gyr of the orbit we released tracer particles from the progenitor, tuning the spatial and kinematic offsets of the escaping stars to best represent the shape of the observed stream close to the progenitor.
The progenitor initially had a stellar mass of $2\times10^8$\,\msun.

The orbit and mock stream are shown in Fig.\ \ref{nbody.fig}.
There are discrepancies on small scales; however,
the model reproduces the overall path, the
higher density of the leading (Eastern) tail, and the asymmetric
broadening of the leading tail where it
curves back toward \gal.

\section{Discussion}
\label{disc.sec}

In this {\em Letter}
we present Dragonfly imaging of the \gal\ system, focusing on its well-known stellar stream.
We find a relatively straightforward system composed of the remnant of a progenitor galaxy, a
leading tail, and a long faint trailing tail. This overall morphology can
be reproduced with a dynamical model without much fine-tuning. In terms of its spatial extent and stellar mass the stream is similar to the Sagittarius stream (see {Sesar} {et~al.} 2017).
The
Milky Way and \gal\ are also quite similar, which means that the entire system offers
an interesting analog to this accretion event.

We now turn to the most puzzling aspect of our study.
As shown in Fig.\ \ref{comparison.fig}
the morphology of the stream in our data is qualitatively different from 
that reported by M08. 
First, we do not confirm the presence of a second (Northern) loop, even though
it contains the brightest part
of the entire M08 tidal stream system.
This stream segment is 
indicated by the $1\arcmin \times 5\arcmin$  yellow box in Fig.\ \ref{comparison.fig}.
From Eq.\ \ref{depth_g.eq} we determine a $3\sigma$ upper
limit of $\mu_g>29.4$\,\ma\ for this region.
Second, the first loop is in a different place:
the location in the Dragonfly image
falls {\em in between} the two loops identified in M08
(see
Fig.\ \ref{comparison.fig}). Other discrepancies are
a greater length of the Western stream
in our data; the presence of a density enhancement
in the first loop (which we identify as the location of the progenitor);
and the much smaller ratio of the apparent width of the stream to the apparent width
of the  \gal\ disk.

It is unlikely that these discrepancies are caused by
a difference in depth or by color variation along the stream.
The M08 image was obtained by an amateur
astronomer in close coordination with professional astronomers,
using a 0.5\,m telescope located on the same site as Dragonfly.
The limiting surface brightness of the M08 image should approach
that of the Dragonfly image when
the size of the telescope, the exposure time (5.8\,hrs in white light and 5.6\,hrs
in red, green, and blue filters), and the
throughput of the filters are taken into account.
Furthermore, neither a difference in depth nor a color gradient
can explain the different locations of the first loop and the other qualitative
discrepancies between the two datasets.
We note that other images of \gal\ in the literature appear to show
only one loop in the same location as in the Dragonfly data
(see Shang et al.\ 1998; Miskolczi et al.\ 2011; Lang, Hogg, \& Scholkopf 2014;
Laine et al.\ 2016). We provide our data on a web page so that others can assess
them.\footnote{See \url{https://www.pietervandokkum.com/ngc5907}.}

There are several routes to make further progress. 
Deeper data can verify the reality
of the tentative sections of the stream and better quantify its substructure.
We will also search for
streams around other galaxies, both in targeted surveys (Merritt
et al.\ 2016; C.\ Gilhuly et al., in preparation) and in
blank field surveys (S.~Danieli et al., in preparation).
More generally, this study follows previous work in demonstrating
the power of the combination of low surface brightness imaging
with dynamical modeling (see also, e.g., {Foster} {et~al.} 2014; {Amorisco}, {Martinez-Delgado}, \&  {Schedler} 2015; {Pearson} {et~al.} 2019).
Systematic surveys of accretion
events across the nearby Universe are providing complementary
information to the extensive work in the Local Group.

\acknowledgements{
We thank Stefan Binnewies, Josef P\"opsel, and Dieter Beer for their help
in understanding their images of
NGC\,5907,
and the referee for insightful comments that improved the manuscript.
}



\end{document}